\documentstyle{article}

\begin{document}

\title{Secret Sharing  Based on a Hard-on-Average Problem}
\date{}
\author{P. Caballero-Gil and  C. Hern\'andez-Goya\\
{\small DEIOC, University of La Laguna, 38271 La Laguna, Tenerife, Spain.} \\
{\small pcaballe@ull.es, mchgoya@ull.es}}
\maketitle
\begin{abstract}

The main goal of this work is to propose the design of secret
sharing schemes based on hard-on-average problems. It includes the
description of a new multiparty protocol  whose main application
is key management in  networks. Its unconditionally perfect
security relies on a discrete mathematics problem classified as
DistNP-Complete under the average-case analysis, the so-called
\emph{Distributional Matrix Representability Problem}. Thanks to
the use of the search version of the mentioned decision problem,
the security of the proposed scheme is  guaranteed. Although
several secret sharing schemes connected with combinatorial
structures may be found in the bibliography, the main contribution
of this work is the proposal of a new secret sharing scheme based
on a hard-on-average problem, which allows to enlarge the set of
tools for designing more secure cryptographic applications.



\end{abstract}

\section{Introduction}\label{intro}

\footnotetext{Research supported by the Spanish Ministry of Education and
Science and the European FEDER Fund under Project
SEG2004-04352-C04-03.\\
Linear Algebra and its Applications. Volume 414, Issues 2-3, 15 April 2006, Pages 626-631 \\
DOI: 10.1016/j.laa.2005.10.040 
}

When a group of nodes in an open network wants to use standard
security services such as authentication, confidentiality or data
integrity, they usually have to share a common secret piece of
information called session key.  In order to build a highly secure
key management service in wireless ad hoc networks, threshold
cryptography based on secret sharing schemes has been yet proposed
in the literature \cite{ZH99}. Since the standard approach of
replication is not suitable for these particularly vulnerable
networks, the distribution of trust among the set of nodes seems
to be the most robust solution. In this work, a new secret sharing
scheme is proposed in order to increase the security level of
known secret sharing schemes by basing them on hard-on-average
problems.

When choosing a specific problem as basis of a cryptographic
application, one of the most important characteristics we have to
look for is that, on the one hand, finding a solution must not be
computationally feasible, whereas on the other hand, generating
pairs formed by (instance, solution) can be efficiently
accomplished. Another usual property of the selected problem is
that the verification procedure for any solution should be as
simple as possible.

The  previous reasons justify the widespread use of problems
belonging to the worst-case $NP$ and $NP - complete$ classes in
the area of the design of cryptographic applications. However,
NP-completeness only guarantees that there is no polynomial-time
algorithm to solve a problem in the worst case. Furthermore, with
the development of the Computational Complexity Theory, and
concretely thanks to the advances made on the Average-Case
analysis, it has been proved that many NP-complete problems may be
efficiently solved when the instance is randomly generated
\cite{Kar76}. So, our proposal to solve this problem is to choose
as base problems those whose difficulty is guaranteed by the
average-case analysis. In this paper, the first (to the knowledge
of the authors) secret sharing scheme based on a problem
catalogued as NP-complete from the point of view of the
average-case analysis is proposed. The problem that has been
chosen as base for the protocol proposed here is the so-called
\emph{Distributional Matrix Representability Problem}
\cite{Wan97}, which is hard-on-average.

The structure of the present work is as follows. The next section
discusses the underlying problem in the framework of average-case
complexity whereas Section 3 locates the proposal in the existing
context of Secret Sharing. The proposed scheme is described in
detail in section 4. Finally, several conclusions and open
problems close the paper.

\section{The Underlying Problem}

The average-case analysis is based on the concept of distributional decision problem
\cite{Lev86}, which is formed by a decision problem and a probability distribution
defined on the set of instances. In this context, the choice of the probability
distribution plays an important role since to fix a probability distribution could
directly influence the practical complexity of the problem. In fact, it has been
proved the existence of NP-complete problems solvable in polynomial time when the
instances are randomly generated under certain distributions.

The distributional class analogous to NP in the hierarchy
associated to the average-case analysis is the DistNP class. It is
formed by pairs containing a decision problem belonging to the NP
class and a  polynomially computable probability distribution. As
in worst-case complexity theory, a distributional problem is said
to be average-case NP-complete (or DistNP-complete) if it is in
DistNP and every distributional problem in DistNP is reducible to
it. The first problem catalogued as DistNP-complete was the
distributional tiling problem, whose formal proof of membership
may been found in \cite{Lev86}. Later, a few works containing the
description of new average-case NP-complete problems have been
published \cite{VR92}, \cite{VL88}, \cite{Gur90},  \cite{Wan97}.

The main difficulty when trying to use problems belonging to this
category in practical applications is the artificiality of their
specifications. So, the principal reason why we have chosen the
Distributional Matrix Representability Problem as base for the
Secret Sharing Scheme here proposed is its naive formulation.

The original Distributional Matrix Representability Problem can be
roughly defined in the following way. All the matrices in the
problem are square and with integer entries. Given an $r \times r$
matrix $A$ and a set of matrices with the same size $\{ A_{1}$,
$A_{2}, ..., A_{k}\}$, it should be decided whether $A$ can be
expressed as a product of matrices in the given set.

In the present work  a bounded version of such a problem for $20
\times 20$ matrices is proposed to be used. In this case the
instance consists of a matrix $A$, a set of $k$ distinct matrices
$\{ A_{1}$, $A_{2}, ..., A_{k}\}$ and a positive integer $n \leq k
$, and the question to answer may be stated as follows: is it
possible to express $A$ as a product of $n$ matrices belonging to
the set $\{ A_{1}$, $A_{2}, ..., A_{k}\}$.
 This bounded version of the
problem was shown to be average-case NP-complete by Venkatesan and Rajagopalan
\cite{VR92}. The distribution considered to generate the integers $k$ and $n$ and the
integer entries of the matrices is the uniform distribution.

The proposed scheme uses the search version of the above
distributional problem, which consists in finding $n$ matrices in
$\{ A_{1}$, $A_{2}, ..., A_{k}\}$ whose product is $A$. The
difficulty of such a version is equivalent to that of the original
distributional decision problem as may be deduced from the general
result stated in \cite{BCGL92}, according to which, search and
decision distributional problems are equivalent from the
average-case analysis point of view.

\section{Secret Sharing Schemes}
Secret sharing multiparty protocols solve usual practical
situations where it is necessary the distribution of a particular
secret $A$ among a set of users $M$. Such a context may be
illustrated with the problem of secret key management. The main
objective of Secret Sharing Schemes is to guarantee that only
pre-designated subsets of participants are able to reconstruct the
secret by collectively combining their shares (or shadows) of $A$.
The specification of all the subsets of participants which are
authorized to recreate the secret is called the access structure
of the Secret Sharing Scheme. An access structure is said to be
monotone if any set which contains a subset that can recover the
secret can itself recover the secret. A general methodology to
design Secret Sharing Schemes for arbitrary monotone access
structure was given in  \cite{c88-Benaloh-Leichter} and
\cite{Ito:Sai:Nis87} . However, such results were not useful for
the Secret Sharing Scheme proposed here because the access
structure is not monotone.

The first secret sharing schemes were independently proposed in
1979 \cite{bla79}, \cite{Sha79}. Later it was demonstrated that
both proposals can be considered described in a common general
scheme due to their basis on the same principles of linear
algebra, \cite{c84-Kothari}. Many different mathematical
structures such as polynomials, geometric configurations, block
designs, Reed-Solomon codes, vector spaces, matroids, complete
multipartite graphs, orthogonal arrays and Latin squares have been
used to model secret sharing schemes. The scheme here described is
based on the generation of a secret matrix as product of matrices
with the same size. Another scheme  based also in matrices was
proposed in \cite{Kar:Gre:Hel83}, but there the secret is a
solution to a system of linear equations.

In general, the structure of most Secret Sharing Schemes is based
on two phases. In the initialization phase, a third trusted party
called the dealer, distributes shares of the secret to authorized
participants through a secure channel. In the reconstruction
phase, the authorized participants of a subset in the access
structure combine their shares to reconstruct the secret. In the
initialization of the Secret Sharing Scheme proposed here the
dealer will publish all the shares and the only secret information
that is revealed to each participant is a pointer to a concrete
share and the names of the other parties in the same access
structure.

Secret sharing schemes which do not reveal any information about
the shared secret to unauthorized individuals are called perfect.
On the other hand, a concept extensively used in Secret Sharing
Schemes is the unconditional security of perfect schemes. A Secret
Sharing Scheme is considered unconditionally secure against
cheaters if the probability of successfully cheating does not
depend on the computational abilities of the cheaters.  The Secret
Sharing Scheme proposed in the following Section is perfect and
unconditionally secure.

\section{A Secret Sharing Scheme Based on an Average-Case Intractable Problem}

In the following secret sharing scheme,  the participation of a
Third Trusted Party is only necessary during the first off-line
initialization stage of the generation of the secret.

Since within the protocol many products of matrices have to be
carried out, it is advisable that all those products are carried
out using the algorithm proposed in \cite{Cop:Win87} due to its
efficiency. On the other hand, the Monte Carlo algorithms
described by Freivalds \cite{Fre79} for the verification of the
product of two matrices may be used to achieve the fraud detection
process. The error probability in these algorithms is bounded by
$2^{-t}$, where $t$ is the number of iterations to be performed.

In order to simplify the general description of the protocol it
will be split into two phases: Initialization and Reconstruction.
\subsection{Initialization}
This set-up stage consists of the generation of the secret, task
that is equivalent to the generation of an instance of the
underlying problem. It should be carried out privately by the
Third Trusted Party.

It starts with the random generation of two integers $k$ and $n$
such that $n \leq k$,
 where $n$ indicates the number of participants.

Then, $k$ $20\times 20$ matrices $A_{i}, i = 1, \ldots, k$ with
integer entries  are randomly generated. All the matrices form a
public set denoted by $M$ and are identified by  indices of their
positions in this set. An
 ordered
 subset of $n$ matrices
constitutes the access structure and their  product, $A$, is
 the secret information.

In this phase  the existence of a secure communication
 channel is necessary because
 each authorized participant has to learn from the Third Trusted Party which is his or her secret shadow in
 $M$, and who is his or her neighbor in the product ordering.
The  main problem of this stage is usually the bandwidth necessary
to transfer the shadows. However, such a difficulty is here
avoided by sending to each authorized participant user only  the
index pointing to the corresponding matrix in the access
structure.

Finally,  the Third Trusted Party sends to the i-th participant a
random binary vector $U_i$ with Hamming weight greater than 1 and
the same dimension that the matrices intervening in the protocol.
The corresponding product vectors between these random vectors and
the secret matrix $A$, $U'_i=A\cdot U_i$, are then published. In
this way, each pair of vectors $U_i$ and $U'_i$ is only available
to  i-th participant, who cannot obtain the secret $A$ from such
information since both vectors define a system of 20 linear
equations in 400 unknowns.

\subsection{Reconstruction} This stage  begins with a verification that allows the following two things: Detect the
presence of cheaters among the shadow holders, and guarantee the
correctness of the secret construction.

In order to allow the verification step, the $n$ different
circular shifts of the original order  of the participants set
$\{P_{(i)}, P_{(i+1)}, \ldots, P_{(i+n-1)}\}$ (where all the
sub-indices are $(mod$ $ (n+1)$) are considered. Such shifts
establish the order in which the verification is developed. The
participant designated in first place according to the shift
($P_{(i)}$) computes
 the product between his or her shadow and the random
binary vector $U_i$ obtaining $A_{(i)}U_i$, then he or she sends
the obtained result to the next participant ($P_{(i+1)}$)
determined by the permutation. In this way, $P_{(i+1)}$ computes
the product determined by his or her shadow and the vector
provided by $P_{(i)}$, $(A_{(i+1)}A_{(i)}U_i)$, and so on. Only if
all the participants have been honest, the last participant
obtains $U'_i$. After comparing it with the published vector, he
or she communicates  the result of the verification to the others.
According to this process, if some participant tries to forge his
or her shadow, the forgery is always detected by the last
participant.

Once passed the verification stage, according to the same
permutations as before, each first participant $P_i$ has to choose
randomly a secret integer $20\times 20$ matrix $X_i$ and to reveal
publicly the product between his or her shadow and $ X_i$ to the
following user.
     The receiver multiplies his or her shadow by the received information
      and sends the result to the following user,
    and
    so on. Finally, the last participant has to give what he or she received to the first user, who will be able to
        recover the secret from $A_{(i-1)}A_{(i-2)}\cdots A_1A_nA_{n-1}\cdots A_i X_i$. In order to do it,
        he or she has to multiply such a matrix by $(A_nA_{n-1}\cdots A_i X_i)^{-1}$ by the right,
        and by
         $(A_nA_{n-1}\cdots A_i X_i)(X_i)^{-1}$   by the left.
\subsection{Comments}

The main parameter of secret sharing schemes is the number of
necessary participants to recover the secret, known as the
cardinality of the privilege users in the access structure. In the
proposed scheme, such a number, denoted by $n$, must be large
enough to avoid that an exhaustive search attack may be
successful. With respect to this point, note that in the original
description of the scheme it is assumed that $n<$20 so that the
participants cannot find $A$ without executing the algorithm. In
general, for any number of participants $n$,  the size of matrices
should be $r>n$ in order to guarantee the security of the scheme.

On the other hand, since the search space for the exhaustive
search is formed by all possible products of $n$ matrices in the
set $M$, and has cardinality $\left(
\begin{array}{c}k+n-1 \\ n
\end{array}\right)$, the number $k$ should also be well chosen.
 The integer $x$ is another important parameter that should be randomly selected in order to
 guarantee the security of the system.

Within the generation of the vector $U$ the binary vectors with
Hamming weight $1$ should be discarded, because otherwise the
vector $U'$ will coincide with a column of the secret matrix $A$.
Thus, the cardinality of the set of possible binary vectors is
$2^{20}-20$.

The security of the proposed scheme is guaranteed by the
difficulty of the distributional matrix representability problem
because any unauthorized individuals only know the public set of
$k$ matrices $\{ A_{1}$, $A_{2}, ..., A_{k}\}$ and a positive
integer $n \leq k $, and from that knowledge they cannot guess
which is the subset that produces the secret $A$. On the other
hand, if the size of matrices is $r>n$, any group of less than $n$
authorized individuals only know a part of the subset that
produces $A$, and have not enough information to find the
remaining matrices.

     The described procedure allows to carry
    out a secure multiparty computation of the secret information without revealing any valuable information about
the shared secret to unauthorized individuals, so the scheme is
perfect. Also, the proposed scheme is unconditionally secure
because the probability of successfully cheating does not depend
on the computational abilities of the cheaters. On the other hand,
the distribution
    of computation  among all the participants is balanced, which turn them into  ideal candidates as primitive tools for
    designing  key management schemes in wireless ad hoc networks.

\section{Conclusions}
In this paper we have proposed a secret sharing scheme based on an
average-case NP-complete problem, the so-called distributional
matrix representability problem. The proposal does not reveal any
valuable  information about the shared secret matrix to
unauthorized parties, and the size of each share equals the size
of the secret, so the scheme is ideal. Also, since the
distribution
    of computation  among all the participants is balanced, the proposal seems a useful primitive for defining  a key management
    protocol in wireless ad hoc networks.    The study of concrete
constructions of difficult instances of the problem that are
adequate according to the design of the scheme is part of a work
in progress.

\bibliographystyle{splncs}

\end{document}